\documentclass[%
preprint,
 amsmath,amssymb,
 aps,
]{revtex4-2}

\usepackage{graphicx}
\usepackage{dcolumn}
\usepackage{bm}
\usepackage{color}

\newcommand{\br}[1]{\left( #1 \right)}

\newcommand{\ex}{\mathrm{e}}

\newcommand{\pa}[1]{\partial_{#1}}

\newcommand{\ovl}[1]{\overline{#1}}

\newcommand{\dd}{\mathrm{d}}

\begin{document}


\title{Spatial-temporal structure functions in Burgers turbulence driven by an Ornstein-Uhlenbeck process}

\author{Jin-Han Xie}
\email{jinhanxie@pku.edu.cn}
\affiliation{Department of Mechanics and Engineering Science at College of Engineering and State key laboratory for turbulence and complex systems, Peking University, Beijing, 100871, PR China}
\affiliation{Joint Laboratory of Marine Hydrodynamics and Ocean Engineering, Pilot National Laboratory for Marine Science and Technology (Qingdao), Shandong 266237, PR China}

\date{\today}

\begin{abstract}
We explore the spatial-temporal structure functions of Burgers turbulence driven by a temporal Ornstein-Uhlenbeck (OU) process, where the characteristic time scale of the OU process is much larger than that of the energy flux across spatial scales. 
Based on the K\'arm\'an-Howarth-Monin equation, we obtain an expression for the third-order spatial-temporal structure function away from the dissipation scale.
This expression combines Kolmogorov's exact result of spatial structure function and the exponential temporal decay of the external force. 
We numerically justify this expression and find that the high-order structure functions also decay exponentially, however, the dependence of decay rates on order is different for the odd- and even-order structure functions.
Comparing the OU-driven Burgers turbulence with that driven by temporal white noise,  their spatial structure functions are identical when the energy injection rates are the same, which justifies Kolmogorov's theory, but these two systems' temporal structure functions differ.
Also, the velocity pdf in the OU-driven Burgers turbulence shows a bimodal distribution, contradicting the near-Gaussian distribution in white-noise-driven turbulence.
\end{abstract}

\maketitle


\section{Introduction}

Turbulence, which involves complicated spatial-temporal information, is ubiquitous in natural and artificial fluid systems.
Structure functions are widely used to describe and study spatial-temporal statistical features of turbulent systems \cite{He2017}.
Most work focuses on spatial structure functions with no temporal difference.
By bridging measurable velocity structure functions with the physically important energy transfer rate across scales, Kolmogorov's structure-function theory establishes a foundation for modern statistical turbulence theory.  
In 1999, \citet{Lindborg1999,Bernard1999,Yakhot1999} developed the exact structure-function theory for two-dimensional turbulence. 
It differs from the theory for 3D turbulence by the coexistence of inverse kinetic and forward enstrophy cascades.
Exact third-order structure function expressions were also derived in other turbulent systems, including Burgers turbulence \cite{E1999,Falkovich2006,Cardy2008,Falkovich2011}, turbulence with bidirectional energy transfer \cite{Alexakis2018,Xie2019,Xie2019b} and anisotropic sheared turbulence \cite{Casciola2003,Wan2009,Wan2010}.  

Exact expressions of structure functions in compressible turbulence are more complicated \cite{Galtier2011,Wang2013,Banerjee2014,Chen2015,Sun2017} because not only velocity structure functions but also density- and pressure-related terms, thus, there coexists multiple modes with distinctive features \cite{Chen2021}. 
High-order structure functions are also important for understanding turbulence statistics, however, in their governing equations the number of unknown independent variables is larger than the number of equations \cite{Hill2001}, leaving the expressions for high-order structure functions not explicitly solvable.

To understand the temporal structure functions, one can identify the characteristic cascade time scale with the characteristic correlation time scale to deduce the temporal correlation based on Kolmogorov's self-similarity assumption \cite{Corrsin1963,Tennekes1972,Monin1975,Landau2010,Podesta2011}. 
Under the assumption that small-scale eddies are randomly swept by large-scale eddies with negligible distortion, from a Lagrangian viewpoint \citet{Kraichnan1964} proposed that the velocity correlation decays as a Gaussian of the temporal difference and the decay rate is scale-dependent.
The random sweeping assumption has been later checked in three-dimensional homogeneous isotropic turbulence \cite{Orszag1972,Chen1989,Sanada1992,Kaneda1999,Chevillard2005,Gorbunova2021}.
Tennekes \cite{Tennekes1975} argued that Kolmogorov's self-similarity assumption applies to Lagrangian velocity.
When mean flow presents, Taylor \cite{Taylor1938} hypothesized that the mean flow carries the spatial structures of turbulent flow without much change, therefore, the spatial and temporal structure functions are linearly related.
Combing the random sweeping and Taylor's frozen hypothesis, \citet{He2006} proposed the elliptic model, which states that the iso-correlation contours in the spatial-temporal difference domain are ellipses.
Detailed information on spatial-temporal correlations in turbulent flows can be found in the review article \cite{He2017}.

Most turbulent systems are driven by temporal white-noise external forcing in the studies of (spatial) temporal structure functions. 
Nevertheless, in realistic scenarios, external forcing has characteristic time scales, such as the periodic tidal forcing in the ocean.
Periodic forcing introduces periodic energy injection rates, and this scenario, named modulated turbulence, has been widely studied \cite{Lohse2000,vonderHeydt2003,vonderHeydt2003b,Cadot2003,Kuczaj2006,Bos2007,Kuczaj2008,Cekli2010}. 

However, the modulated turbulence is not statistically steady, which distinguishes it from statistically steady states in traditional turbulence and brings about extra difficulties in deriving structure-function theories.
So, for simplicity, we consider statistically steady external forcing with prescribed temporal correlation, specifically, a temporal Ornstein-Uhlenbeck (OU) process, whose correlation decays exponentially.
This enables us to study the dependence of structure function in statistically steady states on the external forcing time scale.
The OU forcing was used to drive isotropic turbulence \cite{Eswaran1988,Yeung1989}.
Particularly, \citet{Yeung1989} found that the Lagrangian second-order structure function inherits an exponential dependence on the OU forcing.
Different from their work, this paper focuses on the Eulerian third-order structure function whose spatial dependence is derived from the K\'arm\'an-Howarth-Monin (KHM) equation \cite{Kolmogorov1941,Monin1975,Frisch1995}, but the potential of KHM equation in understanding temporal structure functions is less explored \cite[cf.][]{Hill2006}. 

In this paper, we study Burgers turbulence forced by an OU process for numerical simplicity.  
In \S \ref{sec_theory}, we obtain the expression for the third-order structure-function based on the KHM equation. 
Then we perform numerical simulations to justify the third-order structure function expression and study the high-order structure functions in \S \ref{sec_numeric}. Also, the comparison between the OU-driven Burgers turbulence and that driven by temporal white-noise forcing is presented.
We summarize and discuss our results in \S \ref{sec_diss}.

\section{The third-order structure function} \label{sec_theory}

We start from the forced-dissipative Burgers equation
\begin{equation}\label{burgers_flux}
	u_t + \frac{1}{2}\pa{x} u^2 = F + D,
\end{equation}
where $F$ and $D$ represent external forcing and dissipation, respectively.

In a statistically steady turbulent state, we consider two-point measurements at locations $x$ and $x'=x+r$ with the displacement $r$ and two instances at $t$ and $t'=t+\tau$ with temporal difference $\tau$. By assuming homogeneity, we obtain 
\begin{equation}
	\pa{x} = -\pa{x'} = -\pa{r}. \label{deri}
\end{equation}
Steadiness implies 
\begin{equation}
	\pa{t} = -\pa{t'} = -\pa{\tau}.
\end{equation} 

Multiplying $u'=u(x',t')$ to (\ref{burgers_flux}), adding the conjugate equation and taking an average we obtain
\begin{equation}\label{3rd_eq}
	-\frac{1}{6}\pa{r}\ovl{\delta u^3} = \ovl{u'F} + \ovl{uF'} + \ovl{u'D} + \ovl{uD'},
\end{equation} 
where $\delta u = u'-u$ and the overbar $\ovl{\cdot}$ denotes the average.
Note that the time-derivative term is identically zero in a statistically steady state, and we arrive at a KHM equation that is identical to the one where the two measured points have no temporal difference.

When the external forcing is white-noise in time, considering spatial displacement with $\tau=0$ we have \cite[cf.][]{Srinivasan2012}
\begin{equation}\label{inj_cor}
	\ovl{uF'}+\ovl{u'F} = 2\ovl{F' F},
\end{equation}
and therefore, the spatial third-order structure function becomes
\begin{equation}\label{wn}
	\ovl{\delta u^3}(r,0) = - 12\int_{0}^{r}C(s)\dd s  + \int_{0}^{r} \br{\ovl{u'D} + \ovl{uD'}} \dd r,
\end{equation} 
where $\ovl{F'F} = C(r)$ and $C(0)=\epsilon$ is the energy injection rate.
Considering that energy transfers downscale, ignoring the effect of dissipation away from the dissipation scale we obtain
\begin{equation}\label{3sf_w}
	\ovl{\delta u^3} = - 12\int_{0}^{r}C(s)\dd s,
\end{equation}
which in the limit of $r\to0$ recovers the inertial-range result
\begin{equation}
	\ovl{\delta u^3} = -12\epsilon r.
\end{equation}

When the external forcing $F$ is not a white noise in time, the relation between energy injection and the correlation of the external forcing (\ref{inj_cor}) does not hold, and we may not be able to obtain an explicit expression for the effect of external forcing in the KHM equation.
This is because the temporal integration of the nonlinear term is not guaranteed to be small compared with the effect of forcing.
However, when a temporal scale separation exists between the turbulent nonlinear effect and the forcing's correlation time, the relation (\ref{inj_cor}) may still hold.

To study the turbulence response to a temporally correlated forcing, we consider a simple case where the external forcing $F$ follows an OU process with temporal correlation
\begin{equation}
	\ovl{F' F} = C(r)\ex^{-\sigma |\tau|},
\end{equation}
where $C(r)$ is the externally prescribed forcing correlation and $\sigma$ is the decay rate of the OU process.

But with a temporally correlated forcing, we do not have an exact relation such as (\ref{wn}). So we consider a scenario with temporal scale separation, where the characteristic time scale given by energy flux across spatial scales, $(k_f^2\epsilon)^{-1/3}$, is much smaller than the correlation time scale of the OU process, $1/\sigma$.
Then we conjecture that at a time scale comparable with $1/\sigma$, the third-order structure function inherits the temporal dependence of the OU process.
Thus, based on the knowledge of spatial structure function (\ref{3sf_w}), we obtain
\begin{equation}\label{3sf_general_2}
	\ovl{\delta u^3} = - 12\epsilon\ex^{-\sigma |\tau|}\int_{0}^{r}\frac{C(s)}{{C(0)}}\dd s.
\end{equation} 

Note that when calculating the spatial structure function, based on the downscale energy flux, we can argue that the contribution of small-scale dissipation tends to zero as the viscosity tends to zero \cite{Frisch1995,Bec2000,Xie2018}.
However, for the spatial-temporal structure function, we do not have a similar estimation for the dissipation effect, so we simply assume that with a fixed time scale the dissipation effect also tends to zero as the viscosity tends to zero. 


Particularly, if the OU forcing only acts at one scale ($1/k_f$), i.e., 
\begin{equation}\label{force1}
	F = F_0\br{A \cos(k_fx) + B \sin(k_fx)},
\end{equation}
where $F_0$ is a constant. $A$ and $B$ are independent OU processes generated by
\begin{subequations}
	\begin{align}
		\dd A &= - \sigma A \dd t + \sqrt{2\sigma} \dd W,\\
		\dd B &= - \sigma B \dd t + \sqrt{2\sigma} \dd W,
	\end{align}
\end{subequations}
where $\dd W$ is a white noise with variance one.
Then the correlation of $F$ becomes
\begin{equation}
	\ovl{FF'} = 2F_0^2\sigma \cos(k_fr)\ex^{-\sigma |\tau|}. \label{forcing_correlation}
\end{equation}

Thus, (\ref{3sf_general_2}) becomes
\begin{equation}\label{3rd_sf}
	\ovl{\delta u^3} = - 12\frac{\epsilon}{k_f}  \sin(k_fr)\ex^{-\sigma |\tau|},
\end{equation} 
which we justify in below numerical results.
When $\tau=0$, (\ref{3rd_sf}) recovers the forcing-scale resolving third-order structure-function expression \cite{Bec2000,Xie2019b}; taking a further limit of $kr\ll 1$ the inertial-range result $\ovl{\delta u^3} = - 12\epsilon r$ is obtained \cite{E1997,E1999,E2000,Lindborg2019}.


\section{Numerical results} \label{sec_numeric}
In this section, we run second-order finite-volume numerical simulations for the Burgers turbulence forced by the OU process and white noise with a resolution of $4096$ in a periodic domain of size $2\pi$. 
We add no explicit dissipation, and the injected energy is absorbed by the numerical dissipation by the finite-volume scheme to reach statistically steady states. 
A simulation for Burgers turbulence driven by temporal white-noise forcing is also performed as a comparison.
In these simulations, the Burgers turbulences reach statistically steady states where statistical quantities are calculated.
To obtain the spatial-temporal statistics, we equipartition the interval of the statistically steady state $[T_0,\,T_0+T]$ into $N$ intervals with length $T/N$ and average over the number of intervals $N$.

\subsection{Structure functions in OU-driven Burgers turbulence}

In the OU-driven turbulence, we take $k_f=3$, $\sigma=0.05$ (cf. (\ref{force1}) and (\ref{forcing_correlation})), and $\epsilon\approx420$ is obtained from the statistically steady state by directly calculating the energy injection by external forcing. 
So the time scale separation between the OU process and energy flux across spatial scale is well satisfied with $(k_f^2\epsilon)^{1/3}/\sigma \approx 300 \gg 1$.

\begin{figure}
	\centering
	\includegraphics[width=0.49\linewidth]{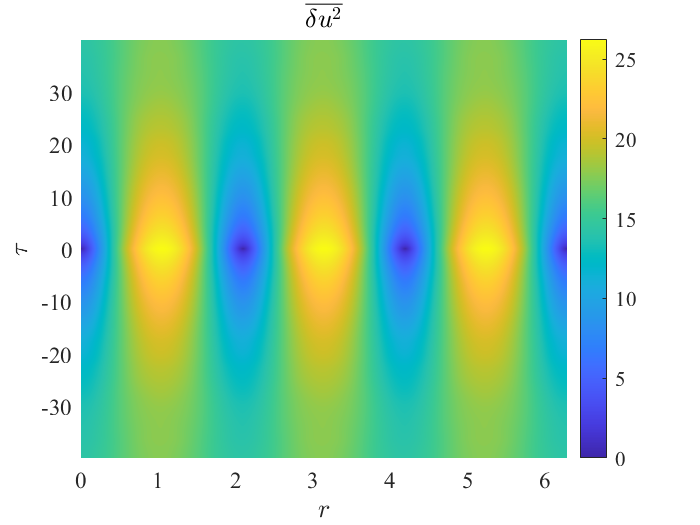}
	\includegraphics[width=0.49\linewidth]{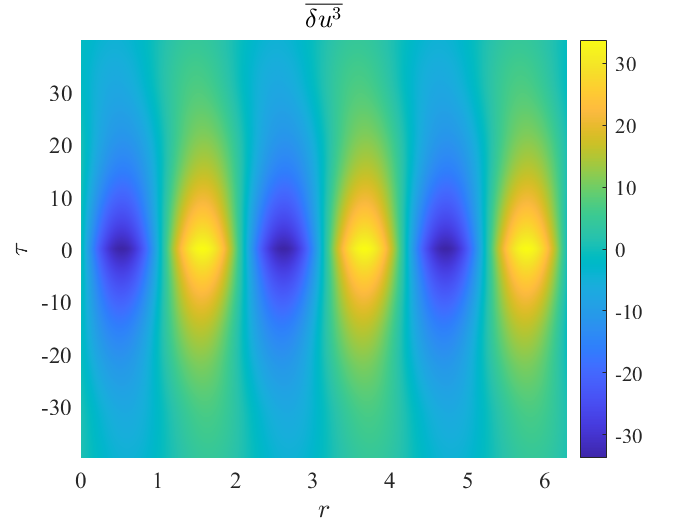}
	\caption{Second- and third-order spatial-temporal structure functions in OU-driven Burgers turbulence.}
	\label{fig_2nd_3nd}
\end{figure}

Figure \ref{fig_2nd_3nd} shows the spatial-temporal second- and third-order structure functions. 
The third-order structure function matches well with the expression (\ref{3rd_sf}), which we check in the following figures with more details.
In this figure, we observe a slight odd signal in $\tau$, i.e., $\tau \to -\tau$ asymmetry, but this asymmetry decreases as the interval used to perform statistics increases, so in the below figures we only show the even parts of the structure functions.

\begin{figure}
	\centering
	\includegraphics[width=0.49\linewidth]{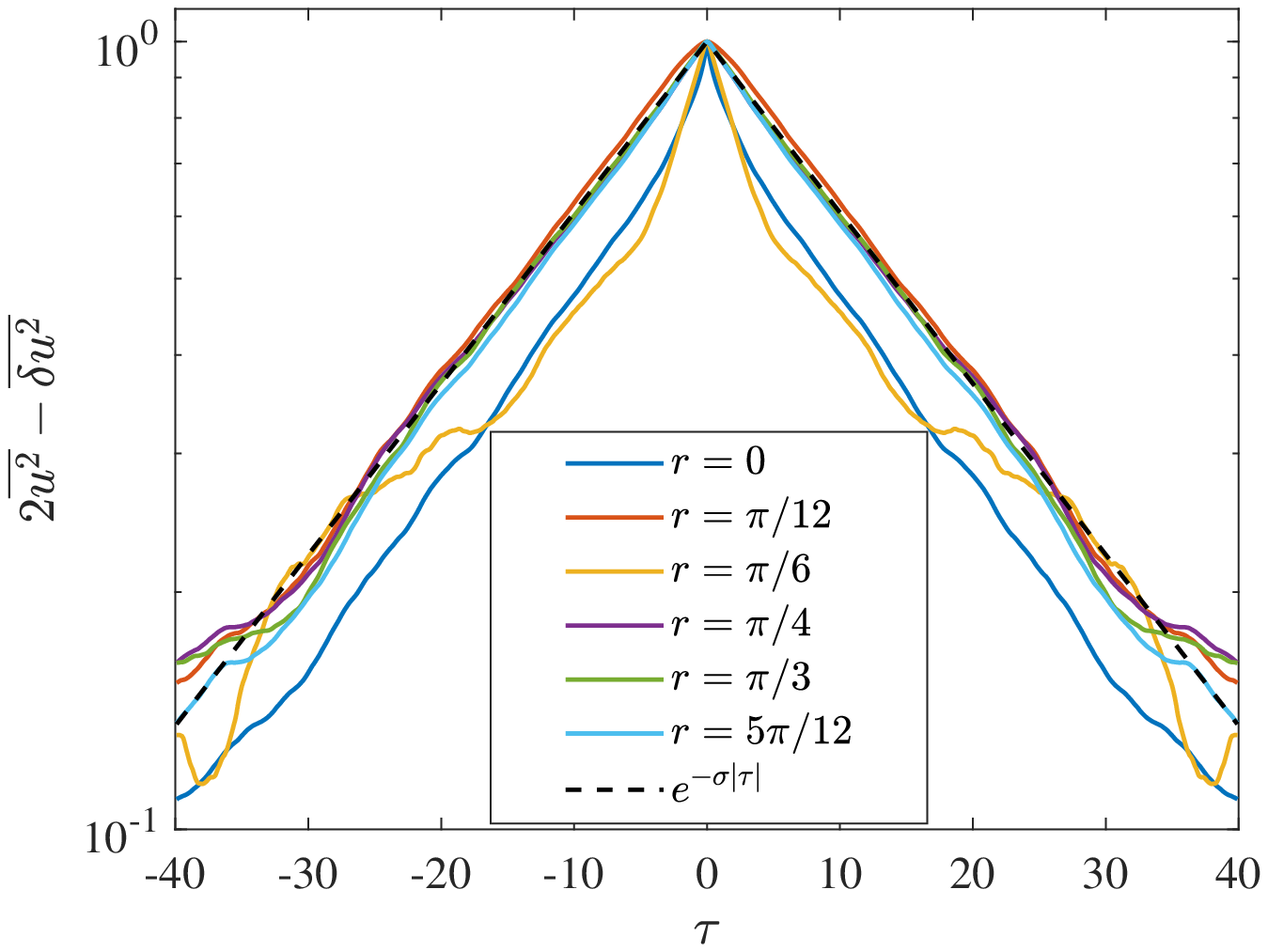}
	\includegraphics[width=0.49\linewidth]{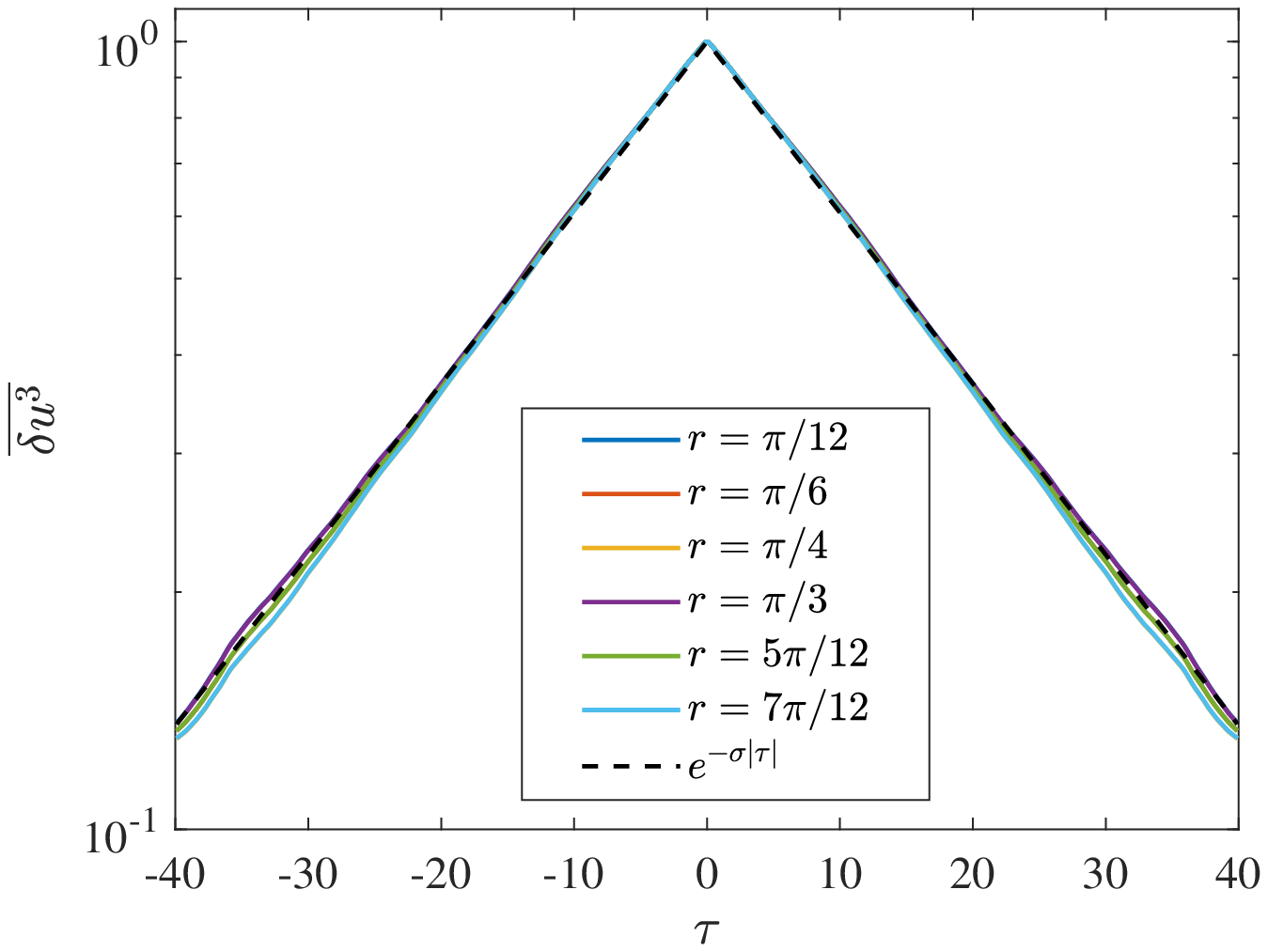}
	\caption{Temporal dependence of the second- and third-order structure functions with different spatial displacements. The structure functions are normalized by their corresponding peak values at $\tau=0$.
		The forcing correlation $\ex^{-\sigma|\tau|}$ is plotted as dashed curves for reference.}
	\label{fig_du2_du3_tau}
\end{figure}

In Figure \ref{fig_du2_du3_tau}, we show the temporal dependence of second- and third-order correlations with fixed spatial displacements.
In the left panel, the $\tau$-dependence of $2\ovl{u'u} = 2\ovl{u^{2}} -\ovl{\delta u^2}$, normalized by its peak value, at $x =0,\, \pi/12,\,2\pi/12,\,3\pi/12,\,4\pi/12$ and $5\pi/12$ are presented.
When $r=\pi/3$, the second-order correlation function reaches its peak value $\ovl{u'u}(\pi/3,0)$, and as the time difference increases, the second-order correlation function decorrelates following an exponential function, $\ex^{-\sigma |\tau|}$, which is inherited from the temporal decorrelation of the external forcing (cf. (\ref{forcing_correlation})).
When $r=\pi/6$, due to the near-zero value of the correlation function (cf. left panel of Figure \ref{fig_2nd_3nd}) the normalized correlation function shows a deviation from the exponential decay.

We present the temporal dependence of third-order structure functions with fixed spatial displacements in the right panel of Figure \ref{fig_du2_du3_tau}.
The fixed spatial displacements are taken as $x = \pi/12,\,2\pi/12,\,3\pi/12,\,5\pi/12,\,6\pi/12$ and $7\pi/12$, and they cover two peak values, at $x=\pi/6$ and $\pi/2$, for the absolute value of the third-order structure function.
We find that these curves match the decorrelation $\ex^{-\sigma |\tau|}$ very well, which justifies our theoretical prediction (\ref{3rd_sf}).

We show the structure functions of different orders with fixed spatial displacements to quantitatively study the temporal correlations. 
Even though in Figure \ref{fig_du2_du3_tau} we show that the second- and third-order structure functions with different displacements have the same decay temporal decay rate, this is not true for higher-order structure functions.
Thus, to capture the temporal dependence brought about by the external forcing, we consider the temporal decay of the structure function with displacement corresponding to the peak of structure functions along $\tau=0$.
Here, we fix the spatial displacements as the ones that take the peak value of the corresponding structure functions along $\tau=0$.
E.g., for the second- and third-order structure functions, the spatial displacements are taken as $r= \pi/3$ and $\pi/4$, respectively.

\begin{figure}
	\centering
	\includegraphics[width=0.49\linewidth]{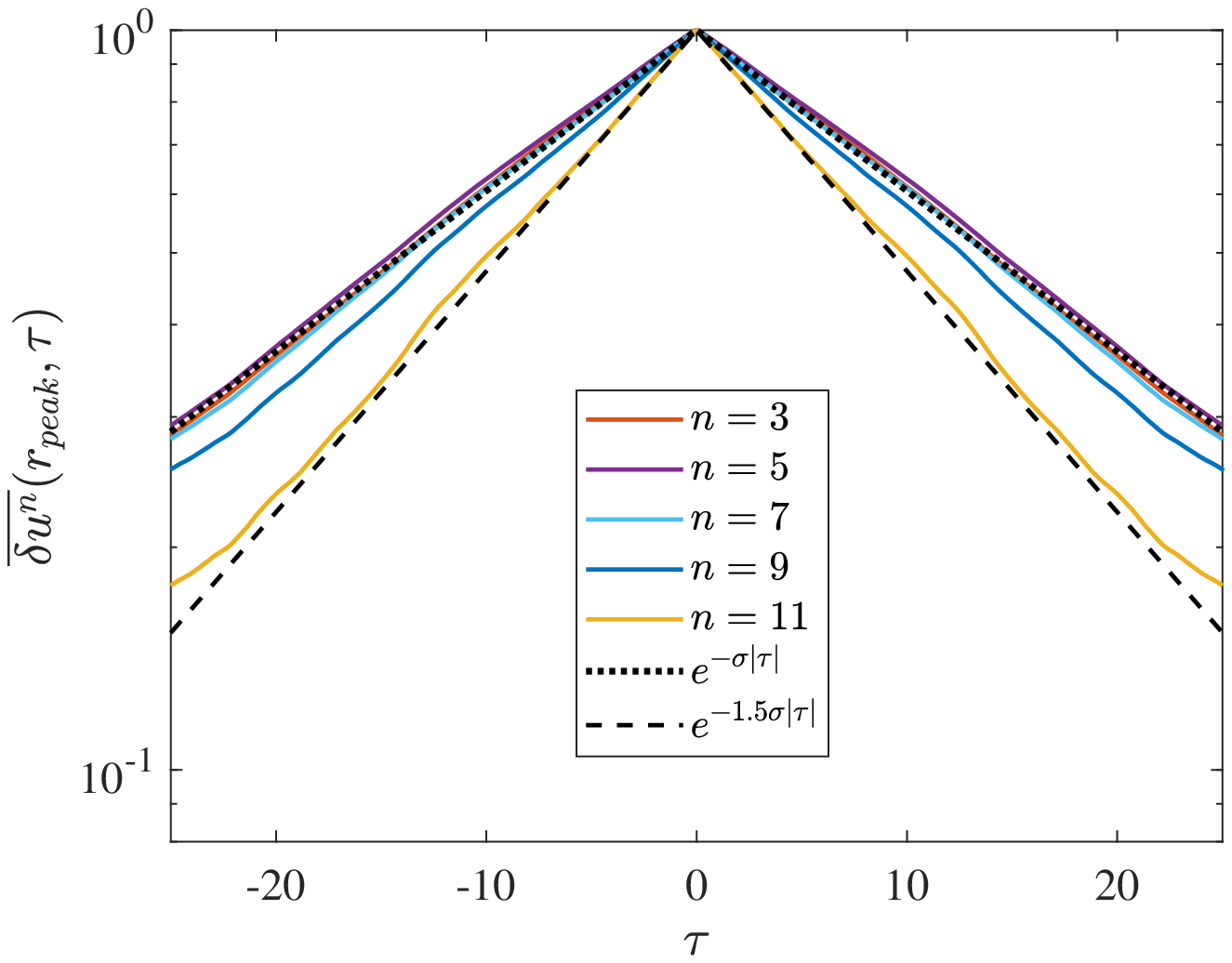}
	\includegraphics[width=0.49\linewidth]{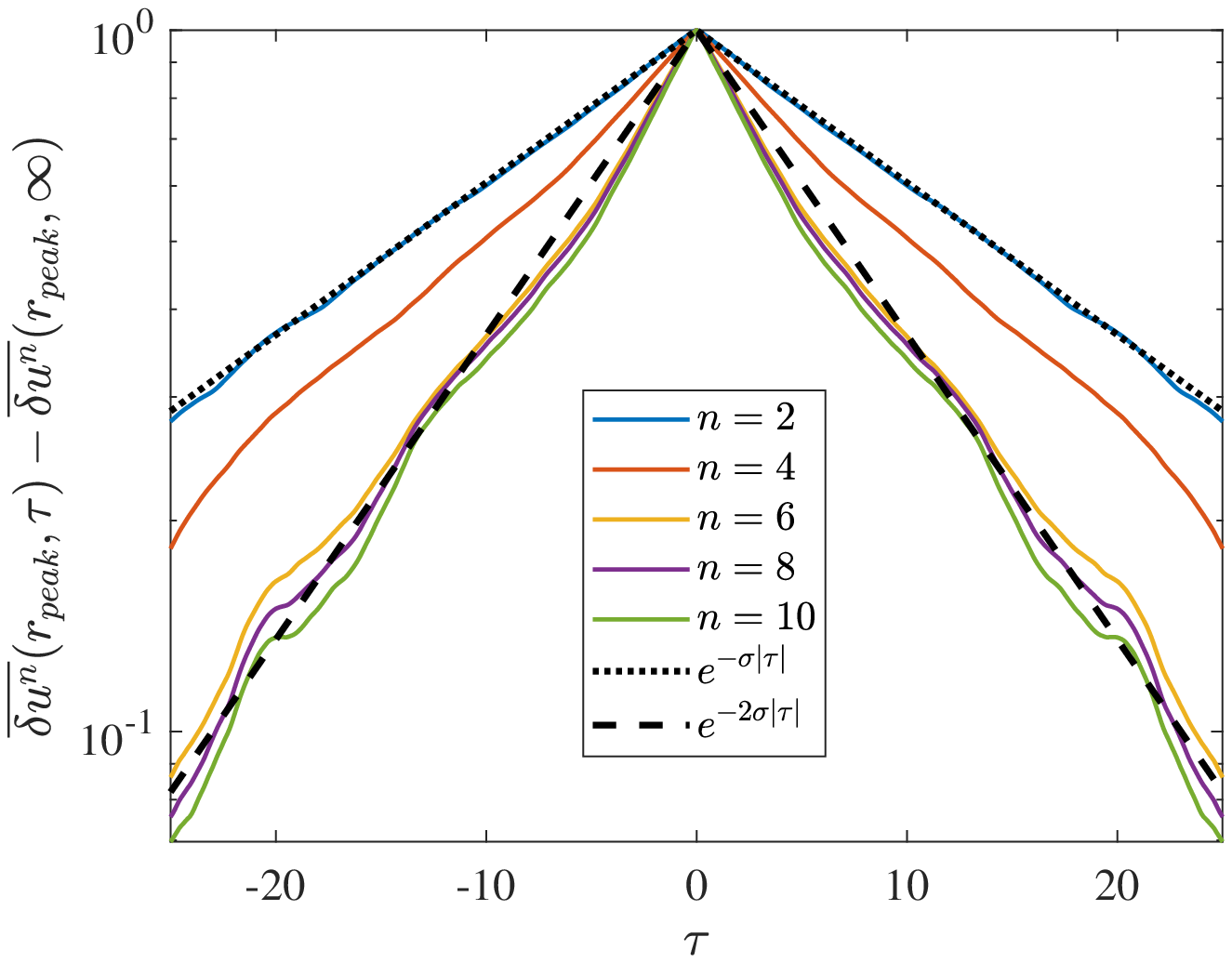}
	\caption{Temporal dependence of  of odd- and even-order structure functions with fixed spatial displacements chosen at each peak along $\tau=0$. Curves with decay rates $\sigma$, $1.5\sigma$ and $2\sigma$ are plotted as references.}
	\label{fig_dun_tau}
\end{figure}

The left panel of Figure \ref{fig_dun_tau} shows that the odd-order structure functions decay exponentially, which resembles the external forcing temporal correlation. For the low orders, $n=3,\,5,\,7$, the decay rate is identical to the decay rate of the external forcing, $\sigma$. 
However, as the order increases, the decay rates increase.
For the even orders,  because the structure functions do not decay to zero as $\tau$ tends to $\infty$, we plot $\ovl{\delta u^{n}}(r_{peak},\tau)-\ovl{\delta u^{n}}(r_{peak},\infty)$ in the right panel of Figure \ref{fig_dun_tau} and find that they also decay exponentially. As the order increases, the decay rates increase and saturate to a decay rate of $2\sigma$.
We can obtain the value of $\ovl{\delta u^{n}}(r_{peak},\infty)$ because the non-zero structure functions at $\tau\to \infty$ are brought about by correlations in the form $\ovl{u^\alpha(x+r,t+\tau)u^\beta(x,t)}$ with $\alpha$ and $\beta$ even integers. 
Considering that $u(x+r,t+\tau)$ and $u(x,t)$ are independent when $\tau\to\infty$, $\lim\limits_{\tau\to\infty}\ovl{u^\alpha(x+r,t+\tau)u^\beta(x,t)} = \ovl{u^\alpha}\ovl{u^\beta}$. 
E.g., $\ovl{\delta u^{4}}(r_{peak},\infty) = 2\ovl{u^4} + 6 \br{\ovl{u^2}}^2$.

We express the temporal decay of structure function with order $n$ as $	\ovl{\delta u^n (r_{peak},\tau)}-\ovl{\delta u^{n}}(r_{peak},\infty) \sim \ex^{-p_n\sigma|\tau|}$, where $p_n$ measures the decay rate. 
The values of $p_n$ are collected in Figure \ref{fig_dun_tau}.

\begin{figure}
	\centering
	\includegraphics[width=0.49\linewidth]{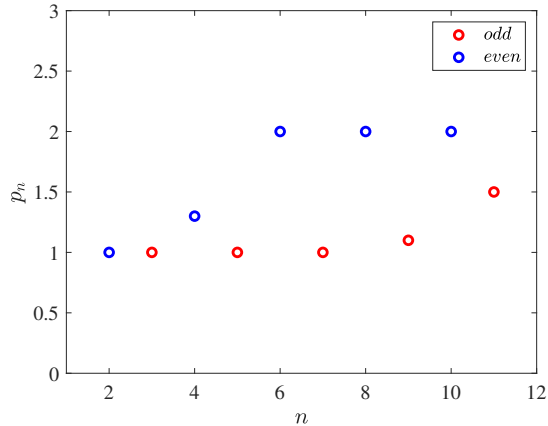}
	\caption{The dependence of the decorrelation rate normalized by $\sigma$ on the order of structure function.}
	\label{fig_p_n}
\end{figure}


\subsection{Comparing white-noise and OU-driven turbulence}

As a comparison, we run a numerical simulation of Burgers turbulence driven by temporal white-noise external forcing, which has the same energy injection rate as that of the OU forcing.

\begin{figure}
	\centering
	\includegraphics[width=0.49\linewidth]{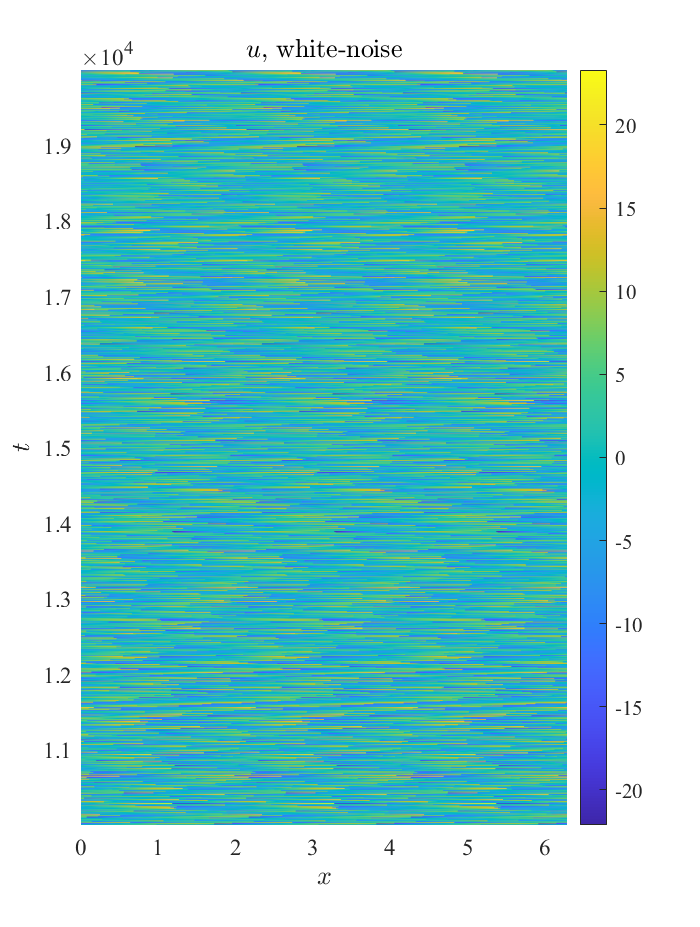}
	\includegraphics[width=0.49\linewidth]{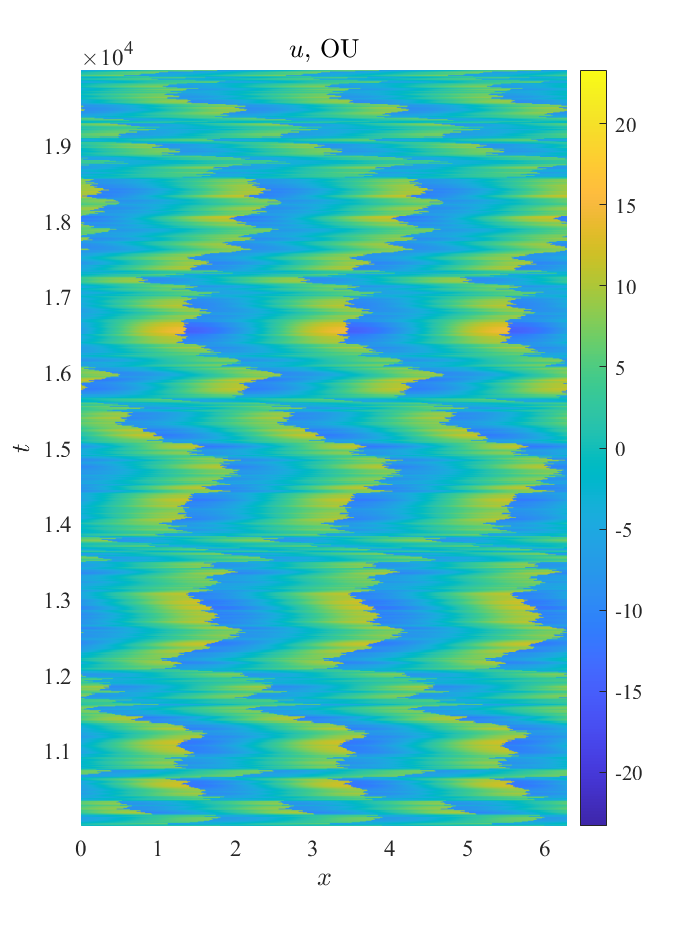}
	\caption{Hovm\"oller diagram for $u$ in Burgers turbulence driven by temporal white-noise and OU forcing, respectively. }
	\label{fig_hovmueller}
\end{figure}

First, we show the Hovm\"oller diagram for field $u$ in two numerical simulations at statistically steady states in Figure \ref{fig_hovmueller}. 
Both panels show large-scale structures corresponding to forcing scale $k_f=3$, while the field driven by OU forcing has a longer time correlation than that forced by temporal white noise.

\begin{figure}
	\centering
	\includegraphics[width=0.49\linewidth]{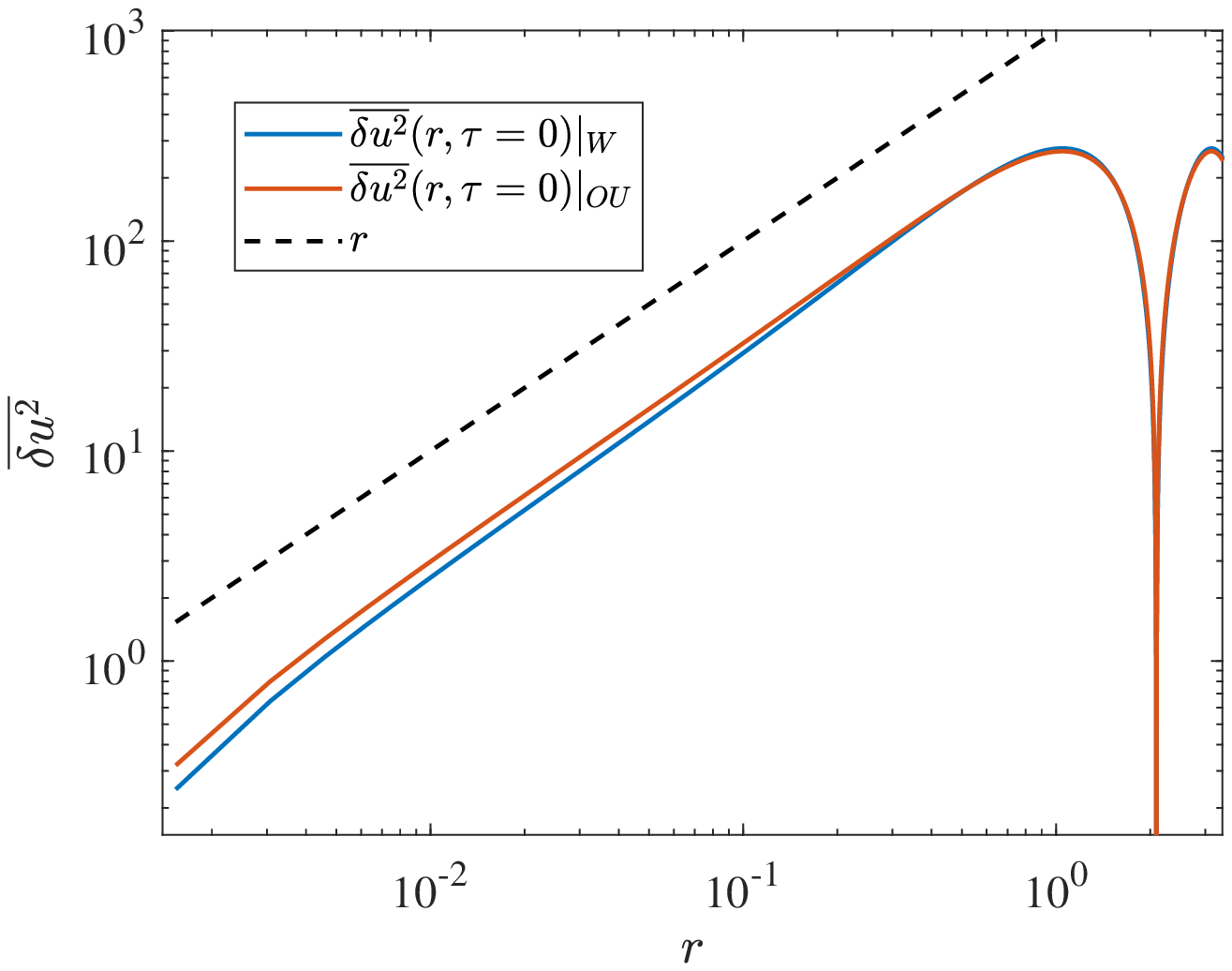}
	\includegraphics[width=0.49\linewidth]{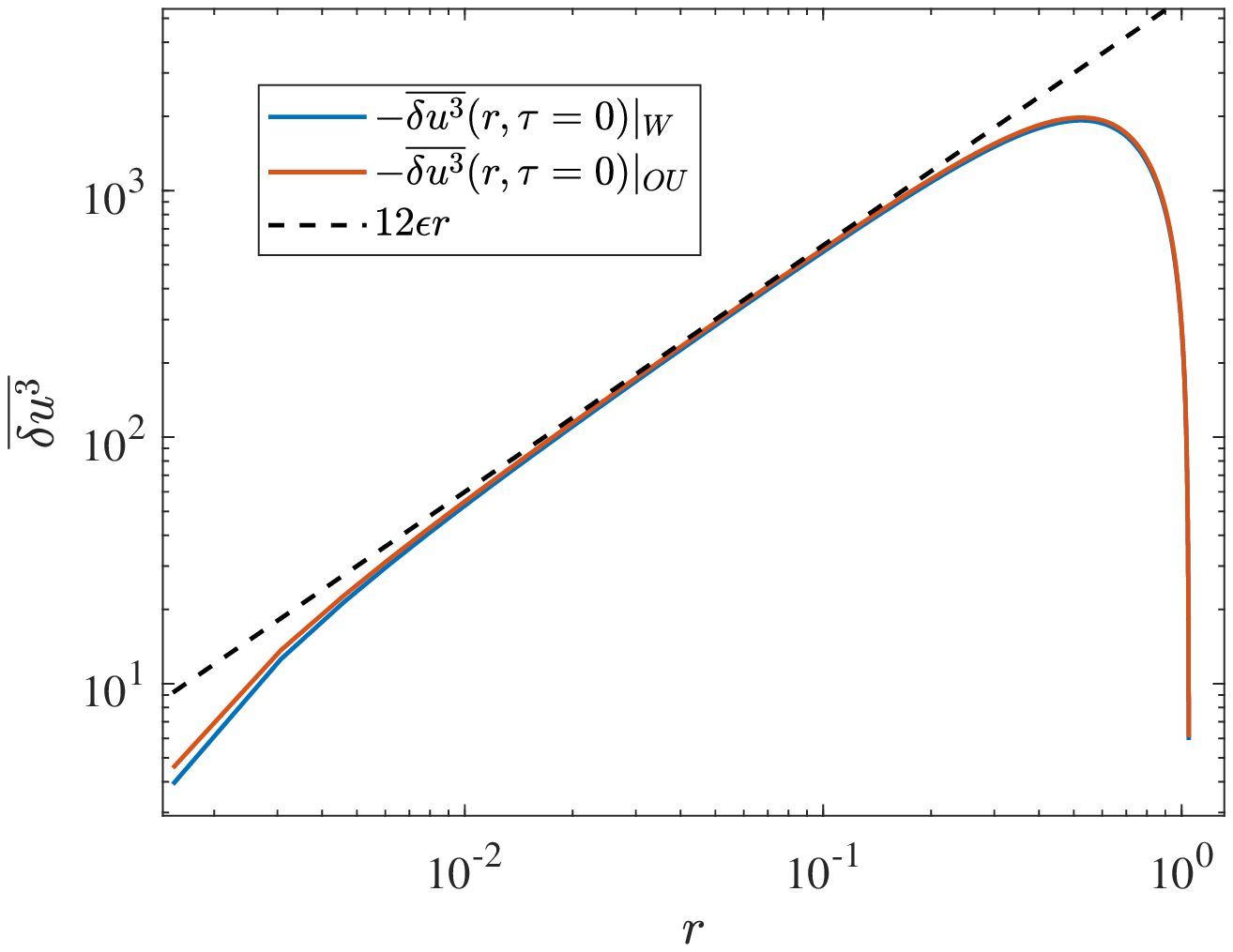}
	\caption{Second- and third-order structure functions with $\tau=0$ in Burgers turbulence driven by temporal white-noise and OU forcing.}
	\label{fig_sf_r_w}
\end{figure}

In Figure \ref{fig_sf_r_w}, we check the second- and third-order spatial structure functions. 
The coincidence of curves reveals the validity of Kolmogorov's scenario in spatial space due to the equal energy injection rate, and we cannot distinguish the two types of turbulence using spatial structure function. 
Also, the $r$ scaling is consistent with the result driven by \cite{E1999}. 

\begin{figure}
	\centering
	\includegraphics[width=0.49\linewidth]{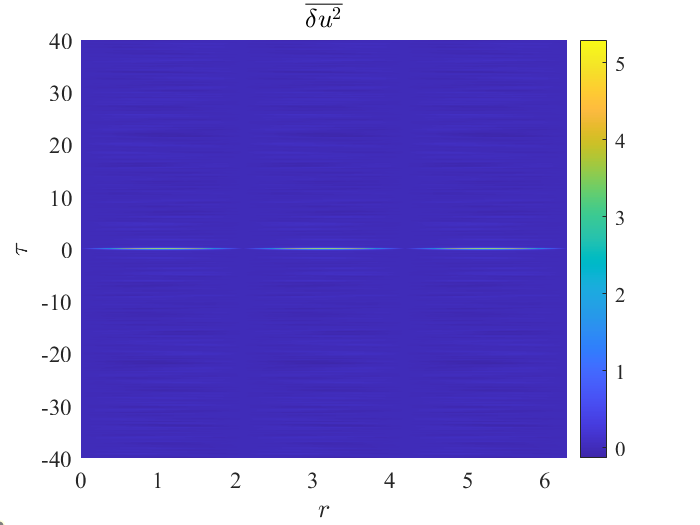}
	\includegraphics[width=0.49\linewidth]{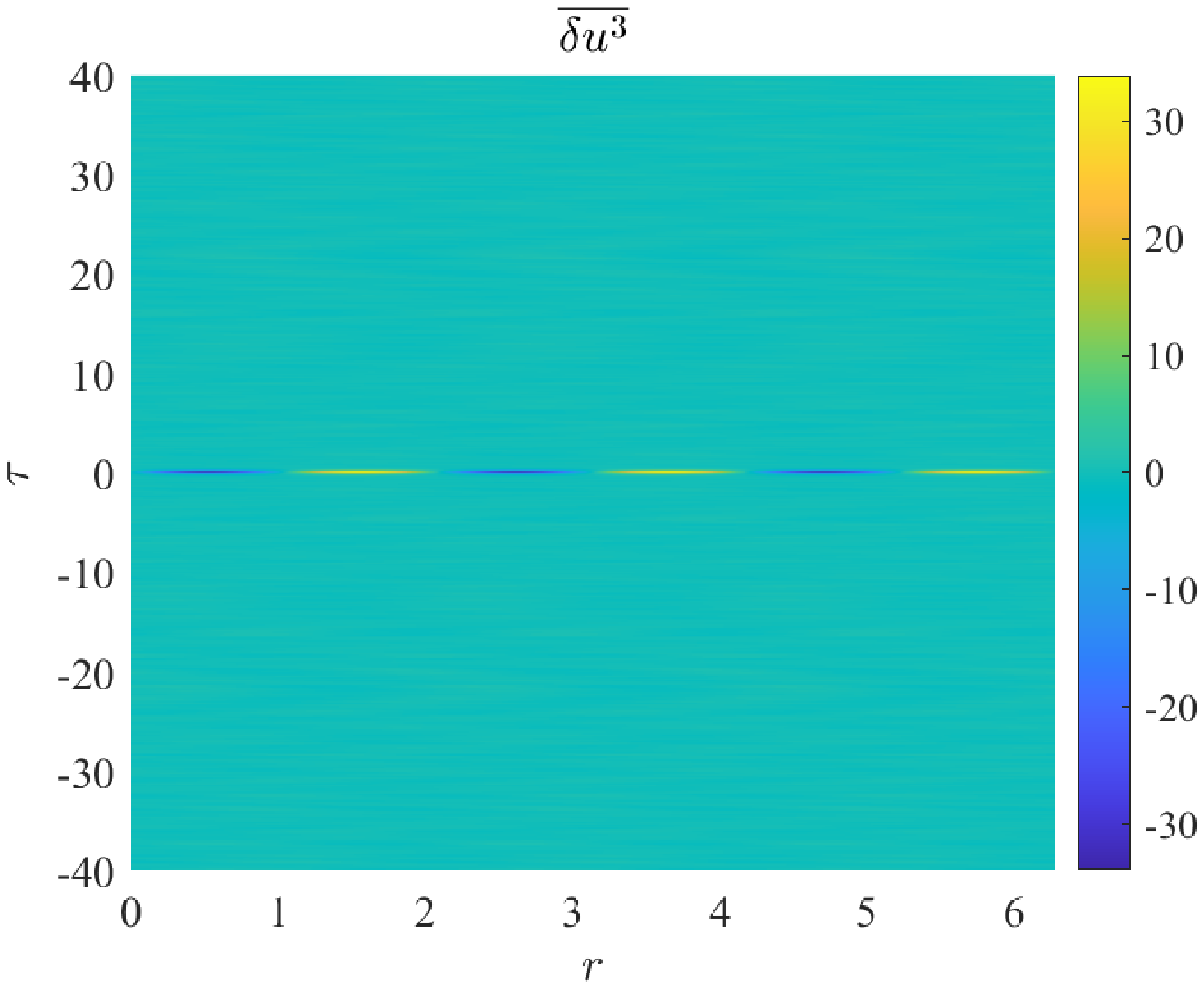}
	\caption{Second- and third-order spatial-temporal structure functions in Burgers turbulence driven by temporal white-noise forcing.}
	\label{fig_2nd_3nd_w}
\end{figure}

\begin{figure}
	\centering
	\includegraphics[width=0.49\linewidth]{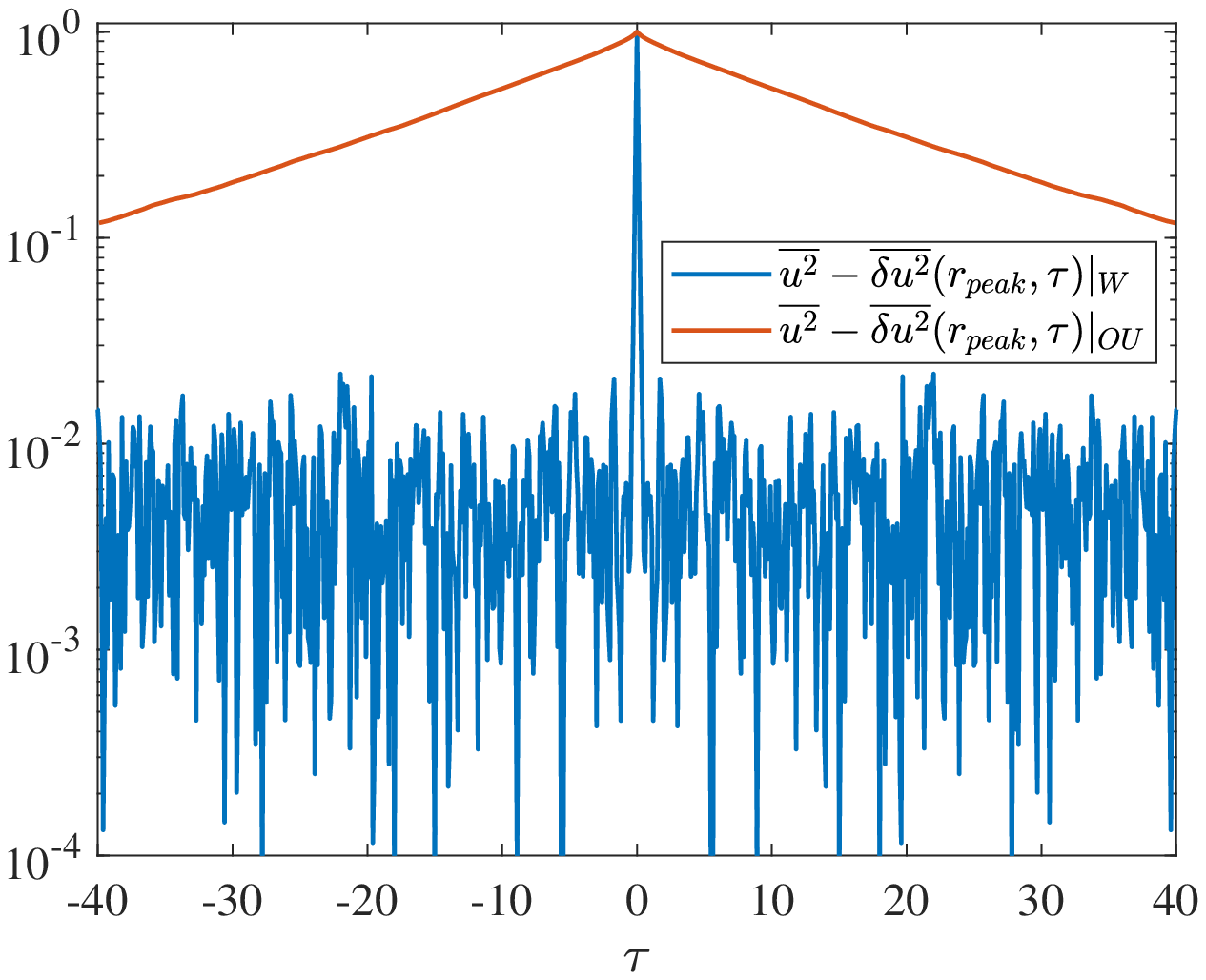}
	\includegraphics[width=0.49\linewidth]{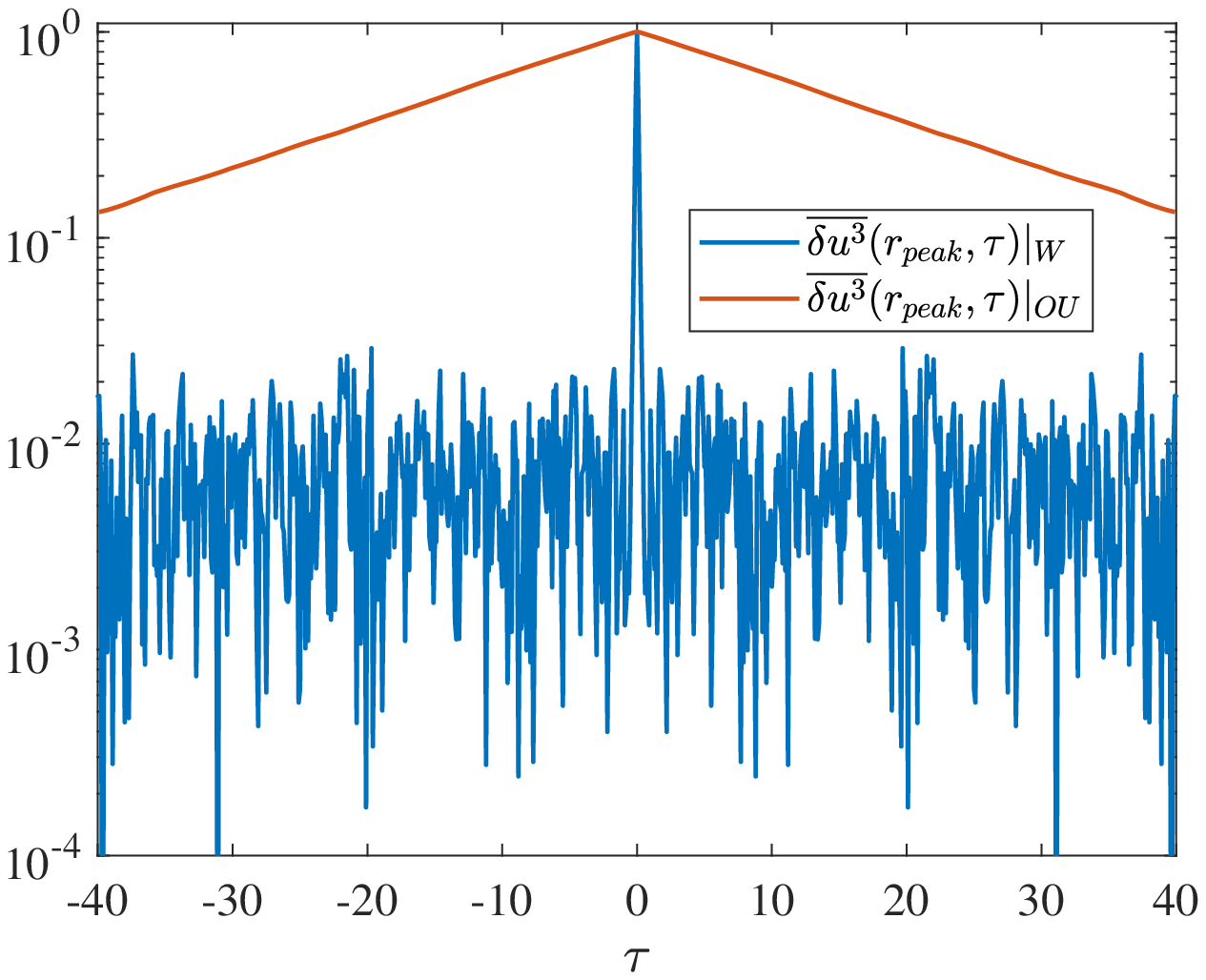}
	\caption{Second- and third-order spatial-temporal structure functions with fixed spatial displacement in Burgers turbulence driven by temporal white-noise forcing.
		The spatial displacement for each structure function is so picked that when $\tau=0$ the structure functions are at the peaks.
		All the structure functions are normalized by their own maximum values taken at $\tau=0$.}
	\label{fig_sf_tau_w}
\end{figure}

The spatial-temporal dependence of the second- and third-order structure functions is shown in Figure \ref{fig_2nd_3nd_w}.
As expected, we observe that structure functions in Burgers turbulence driven by temporal white-noise forcing have much shorter time correlations than those forced by the OU process, which indicates that the long-time correlation in the latter scenario is a result of prescribed forcing.
This shot time correlation is shown in Figure \ref{fig_sf_tau_w} by structures functions with fixed spatial displacement. 

\begin{figure}
	\centering
	\includegraphics[width=0.49\linewidth]{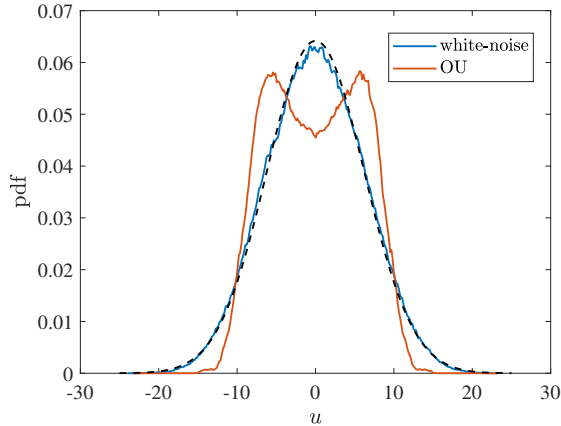}
	\caption{Pdf of $u$ in statistically steady Burgers turbulence driven by temporal white-noise and OU forcing. The black dashed line is a normal distribution for reference.}
	\label{fig_pdf}
\end{figure}

Finally, in Figure \ref{fig_pdf}, we compare the pdf of field $u$ in statistically steady states with two types of forcing.
When driven by temporal white noise, the pdf of $u$ follows a normal distribution, but a bimodal distribution is observed when forced by the OU process.
This bimodal distribution resembles that in the truncated Burgers turbulence \cite{DiLeoni2018}, where the ``tygers" exist \cite[cf.][for the flow structure ``tyger"]{Ray2011}.
However, different from the observation by \cite{DiLeoni2018} where the bimodal distribution is transit, and the final distribution is close to Gaussian, our bimodal distribution is observed in a statistically steady state, which is a result of the OU forcing.
We also need to note that the truncated Burgers system distinguishes from our system, and it remains to study whether there is a link between these two bimodal distributions.

\section{Summary and discussion} \label{sec_diss}

We explore the response of structure functions of Burgers turbulence to the driven by an OU process whose time scale is much larger than that of the energy flux across spatial scales.
Based on the KHM equation, we obtain an expression for the spatial-temporal third-order structure function, which combines the spatial dependence of the classic expression with the exponential temporal decay inherited from the external forcing.
For the structure functions of other orders, we focus on the temporal dependence of structure functions evaluated at the peak for spatial displacement to capture the forcing impact. 
For the odd order, the external forcing's exponential decay rate is inherited by low-order structure function up to order $7$, while for larger orders, the decay rate increases as the order increases.
As to the even order, the second-order structure function decays at the decay rate of the external forcing, and then the decay rate increases and finally saturates at twice the external forcing's decay rate after order $6$. 
Here, since the even order structure function does not decay to zero, the decay rate is extracted from $\ovl{\delta u^{n}}(r_{peak},\tau)-\ovl{\delta u^{n}}(r_{peak},\infty)$. 

In our OU-driven Burgers turbulence, the second-order structure function's exponential decay rate is the same for all spatial scales, making it different from the prediction by random sweeping \cite{Kraichnan1964}, where the decorrelation is Gaussian and scale-dependent. 
This exponential decay agrees with the results calculated from the functional renormalization-group method \cite{Tarpin2018}, but we should distinguish these two results due to different external forcing.
Also, the exponential decay agrees with the numerical discovery by Gorbunova et al. \cite{Gorbunova2021} at a large time difference, but we do not know if there is a link between these two results.

When comparing the Burgers turbulence driven by the OU process with that driven by temporal white-noise forcing with the same energy injection rate, we find that the spatial third-order structure functions are identical for the two systems, justifying Kolmogorov's theory.
As to the pdf of $u$ field, we find that the pdf in the OU-driven system has a symmetric bimodal distribution, contradicting the near-Gaussian distribution for the pdf in the temporal white-noise-driven turbulence.
However, the reason behind this bimodal distribution has yet to be understood.

It needs to be noted that we do not derive the third-order structure-function expression but propose its form and justify it using numerical simulation.
A rigorous derivation starting from stochastic differential equations is still required.

In addition, the exponential decay of the third-order structure function may be linked to the Lagrangian second-order structure function \cite{Yeung2002}, considering that these two statistical quantities may amount to a time derivative.
This link may be obtained from the velocity-acceleration correlation proposed by \cite{Mann1999} and \cite{Ott2000}. 
However, the nonlinear mapping between Lagrangian and Eulerian descriptions must be considered to express the above connection clearly. 

Using the KHM equation to obtain or conjecture a third-order structure-function response to external temporal correlated forcing is potentially to be generalized to other types of forcing. 
This paper only uses the OU process as a heuristic example. 
In our future work, we would follow a similar procedure to study the spatial-temporal structure functions in temporal correlated driven two- and three-dimensional turbulence.\\


\begin{acknowledgments}
J.-H.X. gratefully acknowledges the financial support from the National Natural Science Foundation of China (NSFC) under grant no. 92052102 and 12272006, and Joint Laboratory of Marine Hydrodynamics and Ocean Engineering, Pilot National Laboratory for Marine Science and Technology (Qingdao) under grant No. 2022QNLM010201.
\end{acknowledgments}

%
%


\bibliography{turb_ref}

\end{document}